\documentclass [twocolumn,showpacs,prl]{revtex4}
\usepackage{graphicx}
\usepackage{bm}
\begin{document}
\narrowtext
\title{Nanostructures and enhanced absorption in intense laser interaction with matter: effect of laser prepulses}
\author{P. P. Rajeev\footnote{present address: Steacie Institute for Molecular Sciences, National Research Council of Canada,
Ottawa, ON, K1A 0R6, Rajeev.Pattathil@nrc.gc.ca} S. Kahaly,
S.Bose, P. Prem Kiran, P. Taneja, P. Ayyub and G. Ravindra Kumar
\footnote{electronic address: grk@tifr.res.in}} \affiliation{Tata
Institute of Fundamental Research, 1, Homi Bhabha Road, Mumbai 400
005, India.}
\date{\today}
\newcommand{\e}{\mbox{\boldmath$\eta$}}
\newcommand{\x}{\mbox{\boldmath$x$}}
\newcommand{\si}{\mbox{\boldmath$\xi$}}

\begin{abstract}

Hard x-ray emission (20 - 200 keV) from plasmas produced by
intense laser pulses on nanoparticle coated targets  is compared
with that from optically polished targets. The yield enhancement
offered by nanoparticles is studied under different prepulse
conditions. It is observed that the enhancement reduces when the
nanoparticle coated target is irradiated with  a prepulse with
intensity greater than $10^{13}$ Wcm$^{-2}$. When the prepulse
intensity exceeds $10^{14}$ Wcm$^{-2}$, the enhancement vanishes
completely. This is attributed to preplasma formation on
nanoparticles their subsequent structural modification before the
arrival of the main pulse. It is suggested that high-contrast
ultrashort pulses are essential for nanoparticles to function as
yield enhancers.

\end{abstract}
\maketitle

In recent years, plasmas generated by intense, ultrashort lasers
have attracted multifaceted research to explore  basic physics as
well as  applications. One of the major reasons for this attention
is their high brightness as x-ray sources \cite{Gibbon}. Radiation
and particle emissions from these plasmas are inherently
ultrashort in nature \cite{Murnane}. The radiation pulses are
potentially useful in lithography, time-resolved mapping of
ultrafast atomic and molecular processes, precision imaging
\textit{etc.} \cite{Teubner, Tinten, Westneat}. Since a practical
realization of such sources demands high flux levels, there is a
great deal of interest in methods to enhance the x-ray yield and
the influence of various laser and target conditions has been the
subject of many recent studies. For example, laser prepulses
\cite{Pelletier} and modulated surfaces \cite{Murnane2, Nishikawa,
Kulcsar} enhance laser absorption considerably and subsequently
increase the x-ray yields (mostly in soft and moderately hard
x-ray regimes). Methods of enhancing emission in the very hard
x-ray spectral region are being explored only recently. Such
studies are interesting not only from the point of view of the
enhanced radiation, but also to understand the role of surface
structures or `roughness' in enhancing the production of hot
electrons in plasma responsible for the emission. Enhanced x-ray
yield is a signature of enhanced hot electron production, a
central issue in inertial fusion research \cite{Tabak} and high
energy particle generation and acceleration \cite{Wilks}.

Recently we have reported that metal nanoparticles can be used as
excellent sources of hard x-ray pulses \cite{Rajeev}. A simple,
yet general and quantitative model for the enhanced emission was
also recently presented \cite{Rajeev, OL}. We had proposed that
the non-planar geometry of these nanostructures modifies the local
electric fields around them, resulting in enhanced absorption and
hard x-ray emission. However, to make this method generally
applicable, we need to examine the behaviour of nanoparticles
during their exposure to intense, ultrashort light pulses. Since
intense femtosecond laser pulses invariably possess prepulses as
well as picosecond pedestals - particularly, if sufficient care is
not taken to specifically avoid them - it is important to examine
how nanoparticles respond to prepulses.

The effect of strong prepulses on flat targets and the resulting
particle emissions has been investigated in detail before
\cite{Pelletier}. A prepulse could be intentional or inherent (the
latter could be due to leakage from the cavity dumping in a
regenerative amplifier). Further, the pedestal of an intense pulse
resulting from incomplete pulse compression  or amplified
spontaneous emission, could also serve as a prepulse. If the
prepulse intensity is above the plasma formation threshold, it
would form a plasma layer before the main pulse arrives, thus
crucially altering the interaction process. Depending on the
prepulse-main pulse delay, plasma length-scale could increase and
instabilities can build-up in the plasma, resulting in ripples in
the critical density layer. The unevenness of critical layer as
well as a long density profile favor resonance absorption and
thus, laser absorption and x-ray production increase with preplasma
formation \cite{Pelletier, Sandhu_JAP}. However, the effect of
preplasma formation on nanostructured targets has not yet been
explored. Such a study is very interesting as it can establish the
validity or otherwise of the basic assumptions \cite{Murnane2,
Kulcsar, Rajeev, OL}  in the modelling and understanding of the
phenomenon of enhanced absorption.

In this paper, we address this problem in detail by monitoring
x-ray emission from optically polished copper surfaces as well as
those coated with spherical copper nanoparticles.. We observe that
nanoparticle-coated targets offer 3-4 fold enhancement in hard
x-ray production as compared to uncoated copper targets, when
irradiated with high-contrast  laser pulses (main pulse-prepulse contrast $10^5$:1) at light field
intensities ~ $10^{15} - 10^{16}$ Wcm$^{-2}$.  This enhancement is
examined under different  pre-pulse conditions realized in a
standard two-pulse (prepulse-main pulse) set-up. It is observed
that the enhancement offered by nanoparticles reduces and even
vanishes completely when exposed to prepulses with intensity
levels above a certain threshold. The preplasma formed on
nanoparticles modifies them before the main pulse. This results in
a reduction and complete removal of yield enhancements.  This
study therefore provides crucial information on the exact
mechanism of nanostructure-induced absorption enhancement.
%%%%%%%%%%%%%%%%%%%%%%%%%%%%%%%%%%%%%%%%%%%%%%%%%%%%%%%%%%%%%%%%%%%%%%%%%
\begin{figure}
\centering
\includegraphics [width=3in,height=3in]{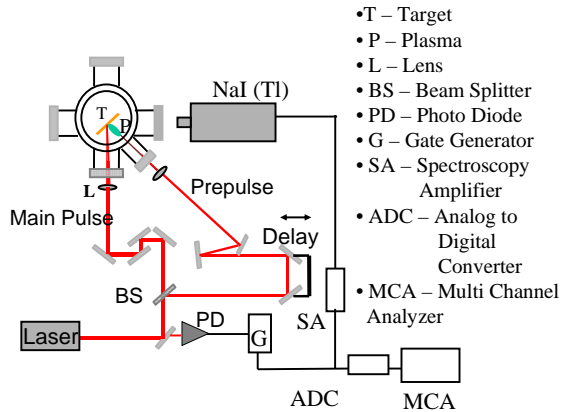}
\caption{Schematics of the Experimental Set-up}
\end{figure}
%%%%%%%%%%%%%%%%%%%%%%%%%%%%%%%%%%%%%%%%%%%%%%%%%%%%%%%%%%%%%%%%%%%%%%%%%

The laser used for our experiments (Fig. 1) is
a custom-designed chirped pulse amplification Ti: Sapphire system,
producing pulses of 100 fs duration at 10 Hz, centered at 806 nm.
The linearly polarized laser beam is split into two and both beams
are focused on the targets housed in a vacuum chamber at $10^{-3}$
Torr, with precise overlap of the focal regions. The target is
continuously translated such that each laser shot irradiates a
fresh area on the target. The weak pre-pulse  and strong main
pulse are focussed respectively by 30 cm and 20 cm focal length
plano-convex lenses. The maximum main beam energy is limited to
4.5 mJ in the present series of experiments, yielding a light
intensity of about 4.5 $\times 10^{15}$ Wcm$^{-2}$ at a focal spot
of 20 microns diameter. A variable optical delay is introduced in the
weaker beam path using a motorized precision translation stage.
The delay between the two beams can be continuously adjusted from
-0.5 ns to +0.5 ns - negative delay implies that  the prepulse
arrives before the main pulse.  We keep the prepulse intensity
sufficiently above the plasma formation threshold ($\sim
10^{13}-10^{14}$ Wcm$^{-2}$ for Copper \cite{APB}) such that a
pre-plasma is formed before the main pulse is incident. The
intensity of the prepulse is varied using a half-wave
plate-polarizer combination. The spatial overlap of main pulse and
prepulse is checked through a CCD imaging system which observes
the plasma formation region. To determine the temporal overlap,
the prepulse reflectivity (at very low prepulse levels, such that
the prepulse does not produce plasma) is monitored as the delay is
changed.  A sharp drop in the prepulse reflectivity, which is
indicative of plasma formation by the main pulse, establishes the
"zero" of temporal overlap \cite{Sandhu_PRL}. X-ray emission from
the plasma is monitored using a time-gated NaI (Tl) scintillation
detector kept in the plane of incidence at $45^{\circ}$ to the
target normal. A detailed description of this diagnostic system
can be found elsewhere \cite{OC}.

%%%%%%%%%%%%%%%%%%%%%%%%%%%%%%%%%%%%%%%%%%%%%%%%%%%%%%%%%%%%%%%%%%%%%%%%%
\begin{figure}
\includegraphics [width=2.8in,height=3in]{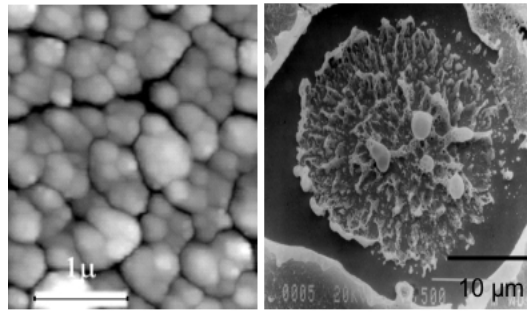}
\caption{Scanning Electron Micrographs of (a) the nanoparticle
coating and (b) destruction of the nanoparticle layer after
exposure to a pulse with intensity over the plasma formation
threshold.}
\end{figure}
%%%%%%%%%%%%%%%%%%%%%%%%%%%%%%%%%%%%%%%%%%%%%%%%%%%%%%%%%%%%%%%

Copper nanoparticles are deposited by high-pressure dc-magnetron
(Atom Tech 320-O) sputtering \cite{Pushan, APB} on optically
polished copper discs held at $0^{\circ}$C. The nanocrystalline
thin films typically consist of a collection of densely packed,
spherical nanoparticles, as shown in Figure 2 (a).  The resulting
nanocrystalline Cu films are optically flat and 1 $\mu$m in
thickness. The crystallographic domain size ($d_{XRD}$) is
obtained from x-ray diffraction line broadening. For a film
deposited in 180 mTorr Ar environment at a sputtering power of
200W, we obtain $d_{XRD}$ = 15nm. Since the thickness of the
nanoparticle layer is greater than the skin depth of the laser
light, the laser essentially interacts only with the film and not
the substrate behind it. However, during the interaction, the film
is locally destroyed in the focal spot, if the laser intensity
exceeds the plasma formation threshold. Figure 2 (b) is a Scanning
Electron Micrograph of one such focal spot. The micrograph, which
represents the surface after  the laser irradiation, further
evidences that the laser interacts just with the nanoparticle
coating leaving the optically flat copper surface behind it
unaffected.

Figure 3 (inset) presents a typical comparison of bremsstrahlung
emission spectrum, measured from an optically polished copper
surface and such a surface coated with spherical nanoparticles,
irradiated at $45^{\circ}$ at $6.0 \times 10^{14}$Wcm$^{2}$ with a
single pulse. The total energy emitted per pulse from a polished
target is $4.2 \times 10^{-14}$ J while the spherical
nanoparticles yield $1.4 \times 10^{-13}$ J, giving about 3 - fold
enhancement in the range 20-200 keV. This is quantitatively
substantiated by a simple model, which ascribes observed yield
enhancements to the electric field enhancements near the
nanostructures \cite{Rajeev, OL}. The model assumes that the
integrity of the nanostructures is preserved during the
interaction time . This assumption is fairly reasonable as the
plasma does not expand significantly to alter the shape of the
structure before the peak of the pulse is reached. Thus, one can
consider the system as a 'nanoplasma', with just a different
dielectric constant; the change in the dielectric function does
not hamper the predictions of the model as the field enhancement
is shown not to  depend critically on dielectric functions, under
our experimental conditions \cite{Rajeev, OL}.

%%%%%%%%%%%%%%%%%%%%%%%%%%%%%%%%%%%%%%%%%%%%%%%%%%%%%%%%%%%%%%%%%%%%%%%%%
\begin{figure}
\includegraphics [width=2.8in,height=3.9in]{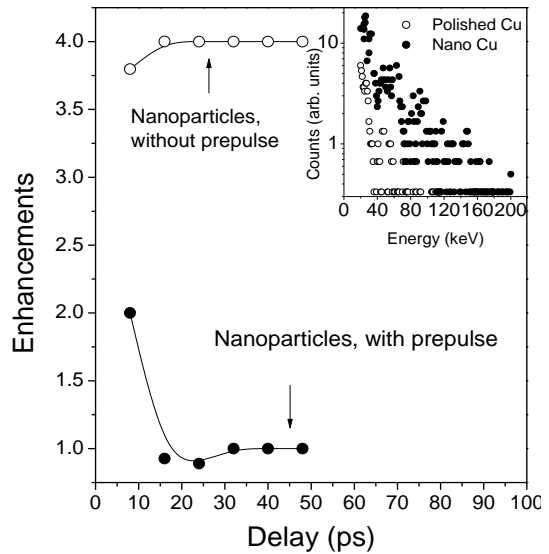}
\caption{Effect of a prepulse on nanoparticle-coated targets at
various delays. The enhancement  factor (ratio of the x-ray yield
of nanoparticle coated targets to that from uncoated targets)
reduces drastically and vanishes with the introduction of a
prepulse at $\sim 10^{14}$ Wcm$^{-2}$. Inset: Typical hard x-ray
spectrum from a polished copper target and a nanoparticle-coated
target}
\end{figure}
%%%%%%%%%%%%%%%%%%%%%%%%%%%%%%%%%%%%%%%%%%%%%%%%%%%%%%%%%%%%%%%%%%%%%%%%%

 We now examine  the effect of prepulses on the nanostructured targets.  Figure 3 shows the result of
 irradiating a spherical nanoparticle-coated target with a prepulse of intensity $\sim 1.5 \times 10^{14}$ Wcm$^{-2}$ at
 normal incidence. The intensity of the main pulse was  around $10^{16}$ Wcm$^{-2}$. A constant
4-fold enhancement was obtained from the
 nanoparticle-coated surface, when irradiated with just the main pulse, without the prepulse, similar to
 the result shown in the inset. This is obviously not a function of the prepulse delay and the plotted data points
 (top curve) indicate just the values obtained with different measurements. As is clear from the figure, there is a
 drastic drop in the enhancement on irradiation of a prepulse, even at small ($\sim$ 10 ps) delays. The enhancement vanishes
 completely as the delay between the prepulse and the main pulse increases.

This observation is very significant as it proves that the
nanostructure has to be intact before the intense pulse, to
observe any yield enhancements. It becomes clear that the
preformed plasma on nanoparticles is not the reason for the
observed enhancements on nanostructured surfaces.  Thus the
observed enhancements in single pulse experiments at these
intensities, themselves substantiate the integrity of the surface
structures in single pulse interaction. Apart from ruling out the prepulse/pedestal levels in
single pulse experiments, this result also suggests that it is
necessary to irradiate with high-contrast pulses for observation
of yield enhancement using nanostructured surfaces.

This is especially important in the context of low contrast
pulses. Typically ultrashort pulses ride over a long pedestal and
wings of duration from picoseconds to nanoseconds. Most high-power
laser systems have prepulses as well. This is mainly caused by the
non-ideal behaviour of the Pockels cell-polarizer combination in
the regenerative amplifier The prepulse just before the cavity
dumping is normally the strongest of all.

Both prepulse and pedestal can form preplasma provided their
intensity levels are above the plasma formation threshold. In flat
targets, as mentioned before, this causes the plasma length-scale
to exceed the wavelength, which is never the case under our
experimental conditions with clean pulses. Further, instabilities
can build-up in a long-lived plasma, resulting in ripples in the
critical density layer. Both of these effects favor resonance
absorption, the major mechanism of light coupling in plasma in our
experimental conditions. However, in the case of nanostructured
surfaces, preplasma formation affects the interaction
detrimentally, as discussed above. The enhanced absorption-
otherwise present- could be reduced or eliminated completely.
%%%%%%%%%%%%%%%%%%%%%%%%%%%%%%%%%%%%%%%%%%%%%%%%%%%%%%%%%%%%%%%%%%%%%%%%%
\begin{figure}
\includegraphics[width=2.8in,height=3.4in]{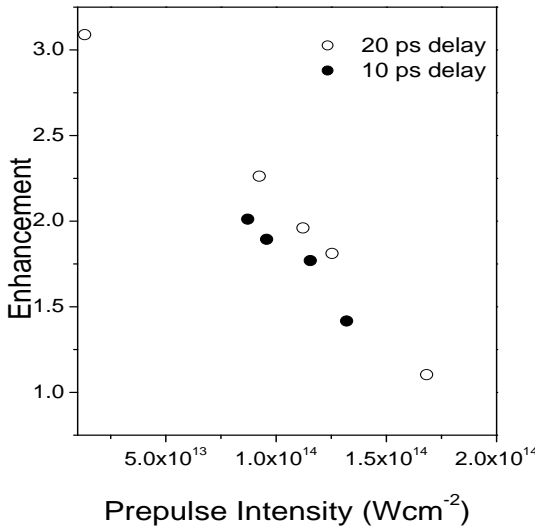}
\caption{Variation of the x-ray yield enhancement with prepulse
intensity and delays. The main pulse intensity is $ \sim 4.5
\times 10^{15}$ Wcm$^{-2}$}
\end{figure}
%%%%%%%%%%%%%%%%%%%%%%%%%%%%%%%%%%%%%%%%%%%%%%%%%%%%%%%%%%%%%%%%%%%%%%%%%

It is thus important to parameterize conditions at which
nanostructures provide enhancements in x-ray yield. We have
studied the x-ray production from laser produced plasma on
nanoparticle-coated targets and flat copper targets under various
prepulse levels. Fig. 4 summarizes the results of such a study.
Yield enhancement from nanoparticles at main pulse intensities
$\sim 4.5 \times 10^{15}$ Wcm$^{-2}$ and under different prepulse
levels is presented. Notice that  even at small delays, the
enhancement starts reducing once the prepulse levels exceed
$10^{13}$ Wcm$^{-2}$. Prepulse levels less than $10^{13}$
Wcm$^{-2}$ do not affect the enhancement. This is understandable
as the nanoparticle coating is about a micron thick and contains
several layers of nanoparticles. Even after some layers are
destroyed or modified as a result of the preplasma formation, some
fresh layers can be preserved for the further interaction if the
prepulse intensity is low enough. Thus, some field enhancement and
subsequent enhanced absorption is still possible with the main
pulse. However, it has to be noted that the enhancement reduces
monotonically with prepulse intensity. Though, the preplasma
formed would be mostly underdense at these intensities, it may not
be produced in a uniform fashion all around the nanostructure. It
is well known that any non-planar geometry severely alters the
local field distribution. It can be seen that most of the field
enhancement is concentrated near the maximum curvature point - at
the tip of the ellipsoidal particle. Thus, plasma is presumably
produced first near the tip of these structures. This spatially
inhomogeneous production of plasma can in turn lead to alterations
in the surface structure, which can reduce enhancement, as
observed. However, this effect may not be strong enough to
completely nullify the enhancements at low prepulse levels.  The
fact that the shape of the critical layer in nanostructures is not
seriously altered even at intensities beyond the plasma formation
threshold is further evidenced in our reflectivity studies
reported elsewhere; the strong dependence of polarization in
absorption was found to be present at intensity levels immediately
beyond  the plasma formation threshold \cite{APB}. It takes a
prepulse intensity $\sim 2 \times 10^{14}$ Wcm$^{-2}$ to
completely nullify the effect of nanoparticles. Lower intensities
do not seem to alter or destroy the nanostructures to an extent
that their effects is removed completely in the interaction with
the following ultrashort pulse.

In conclusion, the hard x-ray emission from plasmas produced on
optically polished copper surfaces and those coated with spherical
copper nanoparticles is examined with femtosecond laser pulses.
The emission is studied under different prepulse levels irradiated
at different times before the main ultrashort pulse. It is
observed that with significant prepulse levels, the enhancement
 in x-ray emission from nanoparticle targets gets adversely affected  and even nullified. These observations
are especially important in studies to find efficient laser plasma sources of
short wavelength radiation. The prepulses and
pedestals invariably associated with high power lasers can affect
the enhanced absorption induced by the nanostructures
detrimentally. Extra pulse cleaning devices such as a Pockels cell
or  saturable absorbers should be used to increase the contrast of
the temporal profile of the pulse in order to use nanostructured
surfaces to facilitate enhanced x-ray production.

\end{document}